\pdfoutput=1
\documentclass[prb,twocolumn,showpacs,amsmath,amssymb,floatfix]{revtex4-1}

\usepackage{graphicx}
\usepackage{dcolumn}
\usepackage{bm}
\usepackage{hyperref}

\begin{document}
	
	\title{Spin-memory loss due to spin-orbit coupling at ferromagnet/heavy-metal interfaces: {\em Ab initio} spin-density matrix approach}
	
	\author{Kapildeb Dolui}
	\affiliation{Department of Physics and Astronomy, University of Delaware, Newark, DE 19716-2570, USA}
	\author{Branislav K. Nikoli\'{c}}
	\email{bnikolic@udel.edu}
	\affiliation{Department of Physics and Astronomy, University of Delaware, Newark, DE 19716-2570, USA}

	\begin{abstract} 
		Spin-memory loss (SML) of electrons traversing ferromagnetic-metal/heavy-metal (FM/HM), FM/normal-metal (FM/NM) and HM/NM interfaces is a fundamental phenomenon that {\em must} be invoked to explain consistently large number of spintronic experiments. However, its strength  extracted by fitting experimental data to phenomenological semiclassical theory, which replaces each interface by a fictitious bulk diffusive layer, is poorly understood from a microscopic quantum framework and/or materials properties. Here we describe an ensemble of flowing spin quantum states using spin-density matrix, so that SML is measured like any decoherence process by the decay of its off-diagonal elements or, equivalently, by the reduction of the magnitude of polarization vector. By combining this framework with density functional theory (DFT), we examine how {\em all three} components of the polarization vector change at Co/Ta, Co/Pt, Co/Cu, Pt/Cu and Pt/Au interfaces embedded within Cu/FM/HM/Cu vertical heterostructures. In addition, we use {\em ab initio} Green's functions to compute spectral functions and spin textures over FM, HM and NM monolayers around these interfaces which quantify interfacial spin-orbit coupling and explain the microscopic origin of SML in long-standing puzzles, such as why it is nonzero at Co/Cu interface; why it is very large at Pt/Cu interface; and why it occurs even in the {\em absence} of disorder, intermixing and magnons at the interface.
	\end{abstract}
	
	\maketitle

Spin-memory loss (SML) of electrons traversing an interface between a ferromagnetic metal (FM) and a normal metal (NM)  affects numerous spin transport phenomena in current-perpendicular-to-plane (CPP) geometry, such as magnetoresistance,~\cite{Bass2007,Dassonneville2010} spin-transfer torque in spin valves~\cite{Grollier2001,Urazhdin2003,Lee2009b} and Josephson current.~\cite{Khasawneh2011} In addition, partial absorption of spin current at an interface between ultrathin layers of $3d$ FM and $5d$ heavy metal (HM) is of paramount importance for understanding  recent spin-orbit torque (SOT)~\cite{Miron2010,Liu2012,Kim2013,Fan2014,Sklenar2016} and spin-pumping-to-charge conversion~\cite{Saitoh2006,Ando2011a,Rojas-Sanchez2014,Jamali2015} experiments---neglecting it  prevents~\cite{Rojas-Sanchez2014} accurate estimation of the spin Hall angle of HM or interfacial enhancement of Gilbert damping of FM.~\cite{Liu2014a} While bulk-SOC-generated effects are well understood,~\cite{Sinova2016} microscopic details of interfacial SOC in FM/HM,~\cite{Moras2015,Grytsyuk2016} as well as FM/NM or HM/NM, bilayers and its effect on spin transport remain largely uncharted.~\cite{Amin2016} For example, theoretical studies~\cite{Haney2013} of SOT often assume simplistic Hamiltonians to describe interfacial SOC and, aside from a handful of studies,~\cite{Mahfouzi2012,Chen2015} standard theory~\cite{Tserkovnyak2005} of spin pumping by precessing magnetization completely neglects interfacial SOC. In general, SML can be expected due to either interfacial SOC or noncollinear magnetic moments at the interface.~\cite{Oparin1999} 

Traditionally, SML  has been quantified~\cite{Bass2007,Dassonneville2010,Khasawneh2011,Rojas-Sanchez2014}  by a phenomenological parameter $\delta$ which determines the probability for electron to flip its spin direction as it traverses the interface. This parameter is typically extracted~\cite{Bass2007,Dassonneville2010} from the measurement of CPP magnetoresistance in spin valves with multilayer insertions by using semiclassical Valet-Fert model~\cite{Valet1993} to express it in terms of the thickness $d$ of a fictitious bulk layer and its spin-diffusion length $\ell_\mathrm{sf}$, $\delta=d/\ell_\mathrm{sf}$. However, processes at the interface---viewed as mathematically sharp plane which is, therefore, shorter than any charge or 
spin dephasing length scale that would make transport semiclassical---require quantum-mechanical description.~\cite{Vedyayev1992,Rychkov2009}  A step in this direction has been undertaken very recently in Ref.~\onlinecite{Belashchenko2016} by expressing $\delta$ in terms of spin-flip transmission and reflection probabilities at the interface, which were also computed via {\em ab initio} calculations (such as for Cu/Pd interface). These probabilities are extracted from the scattering matrix entering the Landauer-B\"{u}ttiker (LB) description of spin-dependent quantum transport~\cite{Rychkov2009,Brataas2006} which views current of charges or spins as a process where they experience transmission and reflection events as they propagate through the system coherently. The magnetoelectronic circuit approach~\cite{Amin2016,Brataas2006}  of Ref.~\cite{Belashchenko2016} introduces tensor conductances that account for both spin precession and loss, but only in the simple case of highly symmetric interfaces it becomes possible to reduce these conductances to reflection and transmission probabilities and to the well-known parameter $\delta$. 

The most general  quantum description of SML can be achieved, akin to any decoherence process,~\cite{Joos2003} by examining {\em how transport across the interface reduces the off-diagonal elements of the spin-density matrix}~\cite{Ballentine2014} 
\begin{equation}\label{eq:rhospin}
\hat{\rho}_\mathrm{spin} = \frac{1}{2} \left( 1+\mathbf{P} \cdot \hat{\bm \sigma} \right),
\end{equation}
The $2 \times 2$ Hermitian matrix $\hat{\rho}_\mathrm{spin}$ of unit trace---as the most general description of spin-$\frac{1}{2}$ states,~\cite{Ballentine2014} which can also be associated with electronic current viewed as an ensemble of ``flowing'' spin quantum states~\cite{Nikolic2005}---is specified by the three real parameters. Therefore, it can equivalently be determined by the polarization vector $\mathbf{P}=\mathrm{Tr}\,[\hat{\rho}_\mathrm{spin} \hat{\bm \sigma}]$, where $\hat{\bm \sigma} = (\hat{\sigma}_x, \hat{\sigma}_y, \hat{\sigma}_z)$ is the vector of the Pauli matrices. Pure spin quantum states, carried by 100\% spin-polarized current, are characterized by $|\mathbf{P}|=1$; incoherent spin states, carried by conventional unpolarized charge current, are characterized by $|\mathbf{P}|=0$; and partially coherent~\cite{Nikolic2005} spin states, carried by partially spin-polarized charge current, are characterized by $0<|\mathbf{P}|<1$. 

\begin{figure}
	\includegraphics[scale=0.48,angle=0]{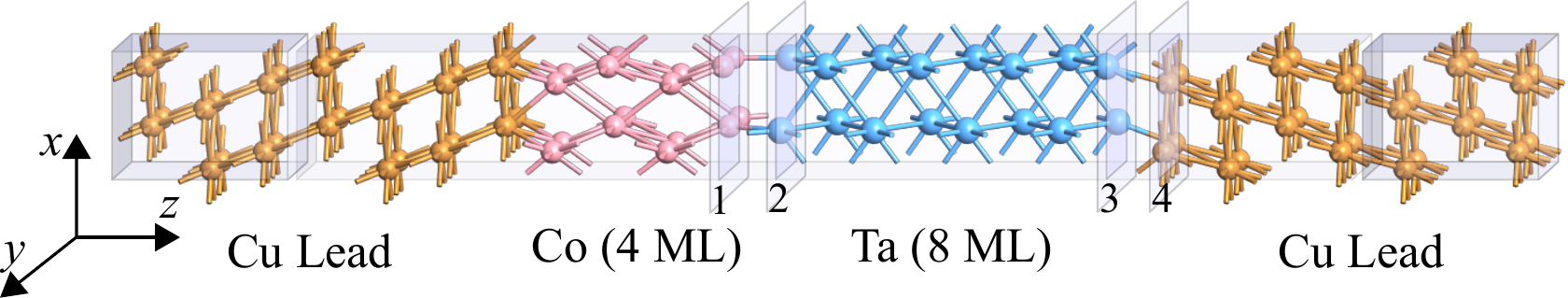}
	\caption{(Color online) Schematic view of a vertical heterostructure for quantum transport calculations of SML, where FM/HM bilayer is sandwiched between two semi-infinite  Cu(111) NM  leads. Small bias voltage $V_b$ applied between the leads injects unpolarized charge current from the left Cu lead along the $z$-axis in the linear-response regime. We consider FM=Co(0001) and HM=Ta(001),Pt(111). The FM layer thickness is fixed at 4 MLs, or it can be semi-infinite (by removing the left Cu lead), while HM layer thickness is varied from 0 to 5 MLs. We calculate spectral functions and spin textures on planes 1 and 2 passing through MLs of FM and HM in direct contact at the FM/HM interface, respectively, or on planes 3 and 4 passing through MLs of HM and Cu around the HM/Cu interface, respectively. The device is assumed to be infinite in the transverse direction, so that the depicted supercell is periodically repeated within the $xy$-plane.}
	\label{fig:fig1}
\end{figure}

In this Rapid Communication, we combine calculation of $\hat{\rho}_\mathrm{spin}$ in terms of nonequilibrium Green functions (NEGF)~\cite{Stefanucci2013} whose input Hamiltonian is obtained from density functional theory (DFT) theory applied to vertical heterostructures  illustrated in Fig.~\ref{fig:fig1} where a bilayer composed of FM=Co(0001) and HM=Ta(001),Pt(111) is attached to one or two Cu(111) semi-infinite leads. The magnetization $\mathbf{m}_\mathrm{Co}$ of the Co layer, which defines direction of ${\bf P}^\mathrm{in}_\mathrm{FM}$ for current impinging onto the interface, is aligned either parallel (i.e., along the $x$-axis in Fig.~\ref{fig:fig1}) or perpendicular (i.e., along the $z$-axis in Fig.~\ref{fig:fig1}) to the interface. The semi-infinite leads are taken into account through the corresponding {\em ab initio} computed self-energies,~\cite{Velev2004} where unpolarized charge current is injected through the left Cu lead and spin-polarized charge current is drained through the right Cu lead. 

The setup in Fig.~\ref{fig:fig1} is inspired by CPP magnetorestance-based experimental measurements~\cite{Bass2007,Dassonneville2010} of $\delta$ where FM/NM bilayer, whose layers are thinner than  $\ell_\mathrm{sf}$, is inserted into Cu layer in between two FM electrodes of a spin valve, except that in our computational scheme we do not need additional right and left FM leads. Our realistic setup  also evades the need for artificial~\cite{Belashchenko2016} introduction of two regions---with and without SOC---within the same HM semi-infinite layer due to fact that spin current or $\hat{\rho}_\mathrm{spin}$ {\em cannot} be uniquely defined along a semi-infinite lead with SOC.~\cite{Nikolic2006}

For clean junctions in Fig.~\ref{fig:fig1} in the ballistic transport regime, we compute $\mathbf{P}^\mathrm{out}_\mathrm{NM}$ for current outflowing into the right Cu lead as a function of the thickness $d_\mathrm{HM}$ of the HM interlayer varied from 0 to 5 MLs. Thus, the parameter~\cite{Rojas-Sanchez2014}
\begin{equation}\label{eq:zeta}
\zeta = \frac{|\mathbf{P}^\mathrm{out}_\mathrm{NM}|}{|\mathbf{P}^\mathrm{in}_\mathrm{FM}|}=\frac{|\mathbf{P}^\mathrm{out}_\mathrm{NM}(d_\mathrm{HM}=1 \  \mathrm{ML})|}{|\mathbf{P}^\mathrm{in}_\mathrm{FM}|}, 
\end{equation}
quantifies SML at FM/HM interfaces, where $|\mathbf{P}^\mathrm{in}_\mathrm{FM}|$ is polarization generated by an infinitely thick FM layer and 
$|\mathbf{P}^\mathrm{out}_\mathrm{NM}(d_\mathrm{HM}=1 \ \mathrm{ML})|$ is polarization of current in NM lead after insertion of 1 ML of HM. In the case of FM layers of finite thickness, we use \mbox{$|\mathbf{P}^\mathrm{in}_\mathrm{FM}| \equiv |\mathbf{P}^\mathrm{out}_\mathrm{NM}(d_\mathrm{HM}=0 \ \mathrm{ML})|$}. Since $\zeta$ measures the fraction of spin current absorbed at the interface, its values in Table~\ref{tab:zeta} can be utilized as an {\em ab initio} boundary condition that is otherwise often introduced~\cite{Rojas-Sanchez2014,Chen2015,Zhang2016b,Amin2016} as a phenomenological  parameter \mbox{$0 \le \zeta \le1$}. Note also that $\zeta=[\cosh(\delta) + r\sinh(\delta)]^{-1}$ can be expressed in terms of measured $\delta$ and spin resistances of different layers which specify parameter $r$ defined in Ref.~\onlinecite{Rojas-Sanchez2014}. Since $\zeta$ can be affected by both FM/HM and HM/NM interfaces in the setup in  Fig.~\ref{fig:fig1}, as also encountered in realistic junctions,  we investigate the effect of the second HM/NM interface by comparing NM=Cu and NM=Au cases. 

By considering current outflowing into the right lead of a two-terminal junction as an ensemble (albeit nonunique~\cite{Ballentine2014}) of  spin quantum states, the expression for $\hat{\rho}_\mathrm{spin}^\mathrm{out}$ was derived in Ref.~\onlinecite{Nikolic2005} in terms of the transmission submatrix $\mathbf{t}$ of the full scattering matrix as the central quantity in the LB approach to quantum transport.~\cite{Rychkov2009,Brataas2006} For the general case of partially spin-polarized current injected from the left lead, whose spins are described by $\hat{\rho}_\mathrm{spin}^\mathrm{in}  = p_\uparrow |\!\! \uparrow\rangle \langle \uparrow \!\!| + p_\downarrow |\!\!\downarrow \rangle \langle \downarrow \!\! |$, 	$\hat{\rho}_\mathrm{spin}^\mathrm{out}$ is given by~\cite{Nikolic2005}
\begin{widetext}
	\begin{equation} \label{eq:rhospint}
	\hat{\rho}_\mathrm{spin}^\mathrm{out} =  \frac{e^2/h}{p_\uparrow (G^{\uparrow \uparrow} + G^{\downarrow \uparrow}) + p_\downarrow(G^{\uparrow \downarrow} +
		G^{\downarrow \downarrow})} \sum_{n^\prime,n=1} 
	\left( \begin{array}{cc}
	p_\uparrow |{\bf t}_{n^\prime n,\uparrow \uparrow}|^2 + p_\downarrow |{\bf t}_{n^\prime n,\uparrow \downarrow}|^2  &
	p_\uparrow {\bf t}_{n^\prime n, \uparrow \uparrow}
	{\bf t}^*_{n^\prime n,\downarrow \uparrow} + p_\downarrow {\bf t}_{n^\prime n,\uparrow \downarrow}
	{\bf t}^*_{n^\prime n,\downarrow \downarrow}  \\
	p_\uparrow {\bf t}^*_{n^\prime n,\uparrow \uparrow} {\bf
		t}_{n^\prime n,\downarrow \uparrow} + p_\downarrow {\bf t}^*_{n^\prime n,\uparrow \downarrow} {\bf t}_{n^\prime n,\downarrow \downarrow} &
	p_\uparrow |{\bf t}_{n^\prime n,\downarrow\uparrow}|^2 + p_\downarrow |{\bf t}_{n^\prime n,\downarrow \downarrow}|^2
	\end{array} \right).
	\end{equation}
\end{widetext}
Here ${\bf t}_{n^\prime n,\sigma^\prime \sigma}$ are complex numbers determining probability amplitude for a spin-$\sigma$ electron incoming from the left lead in the orbital state $|n \rangle$ to appear as a spin-$\sigma^\prime$ electron in the orbital channel $|n^\prime \rangle$ in the right lead. The zero temperature spin-resolved conductances $G^{\sigma\sigma'}$ are given by the LB formula, \mbox{$G^{\sigma\sigma'} = \frac{e^2}{h} \, \sum_{n^\prime,n=1} |{\bf t}_{n^\prime n, \sigma \sigma'}|^2$}. Using 100\% spin-polarized current injection, $p_\uparrow=1.0$ and $p_\downarrow=0$, Eq.~\eqref{eq:rhospint} can be used to study SML at NM/HM interfaces within NM/HM/NM junctions.~\cite{Belashchenko2016} For unpolarized current injection,  $p_\uparrow=p_\downarrow=0.5$, where current can be subsequently polarized by explicitly introducing FM layer as in Fig.~\ref{fig:fig1}, polarization vector can be equivalently calculated~\cite{Chang2014a} using $P_\alpha = \mathrm{Tr}[\hat{\sigma}_\alpha \mathbf{t} \mathbf{t}^\dagger]/\mathrm{Tr}[\mathbf{t} \mathbf{t}^\dagger]$. We obtain  the transmission matrix, \mbox{$\mathbf{t}=\sqrt{-2\mathrm{Im}\, {\bm \Sigma}_{\mathbf{k}_\parallel}^{R}(E)} \cdot \mathbf{G}_{\mathbf{k}_\parallel}(E) \cdot \sqrt{-2\mathrm{Im}\, {\bm \Sigma}_{\mathbf{k}_\parallel}^{L}(E)}$}, using the retarded Green's function (GF)
\begin{equation}\label{eq:rgf}
\mathbf{G}_{\mathbf{k}_\parallel}(E)= [E-\mathbf{H}^\mathrm{DFT}_{\mathbf{k}_\parallel} - {\bm \Sigma}_{\mathbf{k}_\parallel}^\mathrm{L}(E) - {\bm \Sigma}_{\mathbf{k}_\parallel}^\mathrm{R}(E)]^{-1}, 
\end{equation}
where $\mathbf{H}^\mathrm{DFT}_{\mathbf{k}_\parallel}$ is DFT Hamiltonian computed for the active region composed of FM/HM bilayer with 4 MLs of the leads attached on each side,  $\mathbf{k}_\parallel=(k_x,k_y)$ is the transverse $k$-vector, and ${\bm \Sigma}_{\mathbf{k}_\parallel}^\mathrm{L,R}(E)$ are the self-energies~\cite{Velev2004} due to the left (L) and the right (R) semi-infinite leads. 

\begin{figure}
	\includegraphics[scale=0.36]{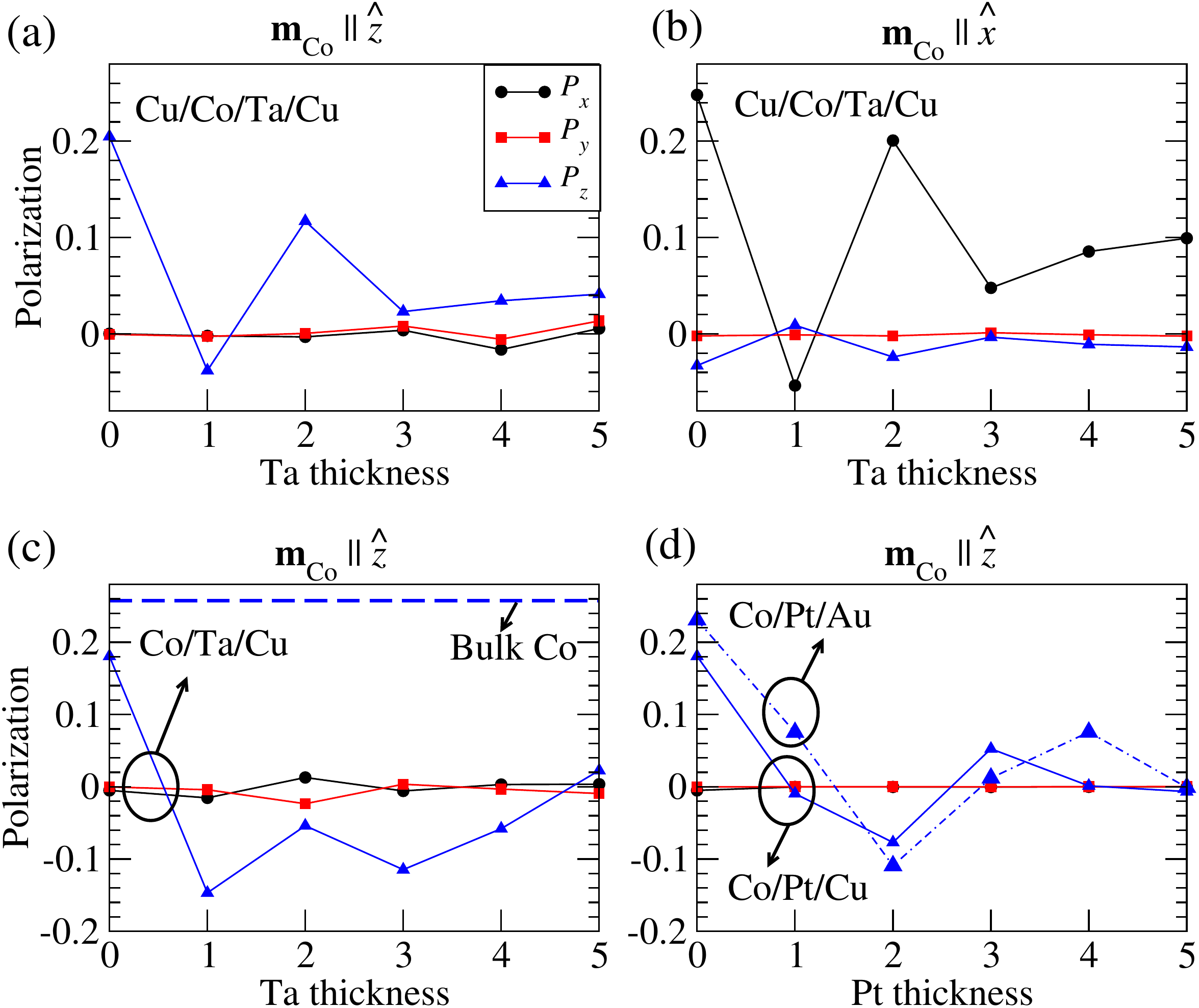}
	\caption{The components $(P_x,P_y,P_z)$ of the polarization vector $\mathbf{P}$, defined by Eq.~\eqref{eq:rhospin}, as a function of the thickness of HM=Ta,Pt layer within vertical heterostructures: (a) Cu/Co(4 ML)/Ta(n ML)/Cu with magnetization $\mathbf{m}_{\rm Co} || \hat z$ pointing along the direction of transport  and perpendicular to the interface; (b) Cu/Co(4 ML)/Ta(n ML)/Cu with magnetization $\mathbf{m}_{\rm Co} || \hat x$ parallel to the interface;  (c) Co/Ta(n ML)/Cu; and (d) Co/Pt(n ML)/Cu or Co/Pt(n ML)/Au (in the latter case, short dashed line plots only $P_z$). The horizontal dashed line in panel (c) denotes polarization $|\mathbf{P}|=P_z \simeq 0.26$ of current flowing through infinite homogeneous Co(0001) layer with its magnetization parallel to the direction of transport.}                   
	\label{fig:fig2}
\end{figure}

Prior to DFT calculations of $\mathbf{H}^\mathrm{DFT}_{\mathbf{k}_\parallel}$, we employ the interface builder in the VNL package~\cite{vnl} to construct a common unit cell for bilayers. In order to determine the interlayer distance and relaxed atomic coordinates, we perform DFT calculations using VASP package~\cite{vasp,Kresse1993} with  Perdew-Burke-Ernzerhof (PBE) parametrization of the generalized gradient approximation for the exchange-correlation (XC) functional and projected augmented-wave~\cite{Blochl1994} description of electron-core interactions. The cutoff energy for the plane wave basis set is chosen as 600 eV, while $k$-points were sampled at $9 \times 9$ surface mesh. The atomic coordinates are allowed to relax until the forces on each atom are less than $0.01$ eV/\AA{}, while keeping the interlayer distances fixed. {\em Ab initio} quantum transport calculations are performed using ATK package~\cite{atk} where we use PBE XC functional, norm-conserving pseudopotentials for describing electron-core interactions, and SG15 (medium) type local orbital basis set.~\cite{Schlipf2015} The energy mesh cutoff for the real-space grid is chosen as 76.0 Hartree. Transmission matrices are obtained by integrating over 251$\times$251 $k$-point mesh. 

\begin{table}[t]
	\caption{Parameter $\zeta$, defined in Eq.~\eqref{eq:zeta} to quantify SML at interfaces, for different types of junctions studied in Fig.~\ref{fig:fig2}. The magnetization of the Co layer is oriented along the $z$-axis perpendicular to the interface, except for the value in parentheses where $\mathbf{m}_\mathrm{Co} \parallel \hat{x}$ is in the plane of the interface.}\label{tab:zeta}
	\begin{tabular}{ccccccc}
		\hline 
		& Cu/Co/Ta/Cu  &  Co/Ta/Cu  & Co/Pt/Cu  & Co/Pt/Au & Co/Cu & Co/Au \\
		\hline \hline
		$\zeta$  & 0.19 (0.22) &  0.57 & 0.04 & 0.29 & 0.70 & 0.90 \\ 
		\hline 
	\end{tabular}
\end{table} 

\begin{figure*}
	\includegraphics[scale=0.40]{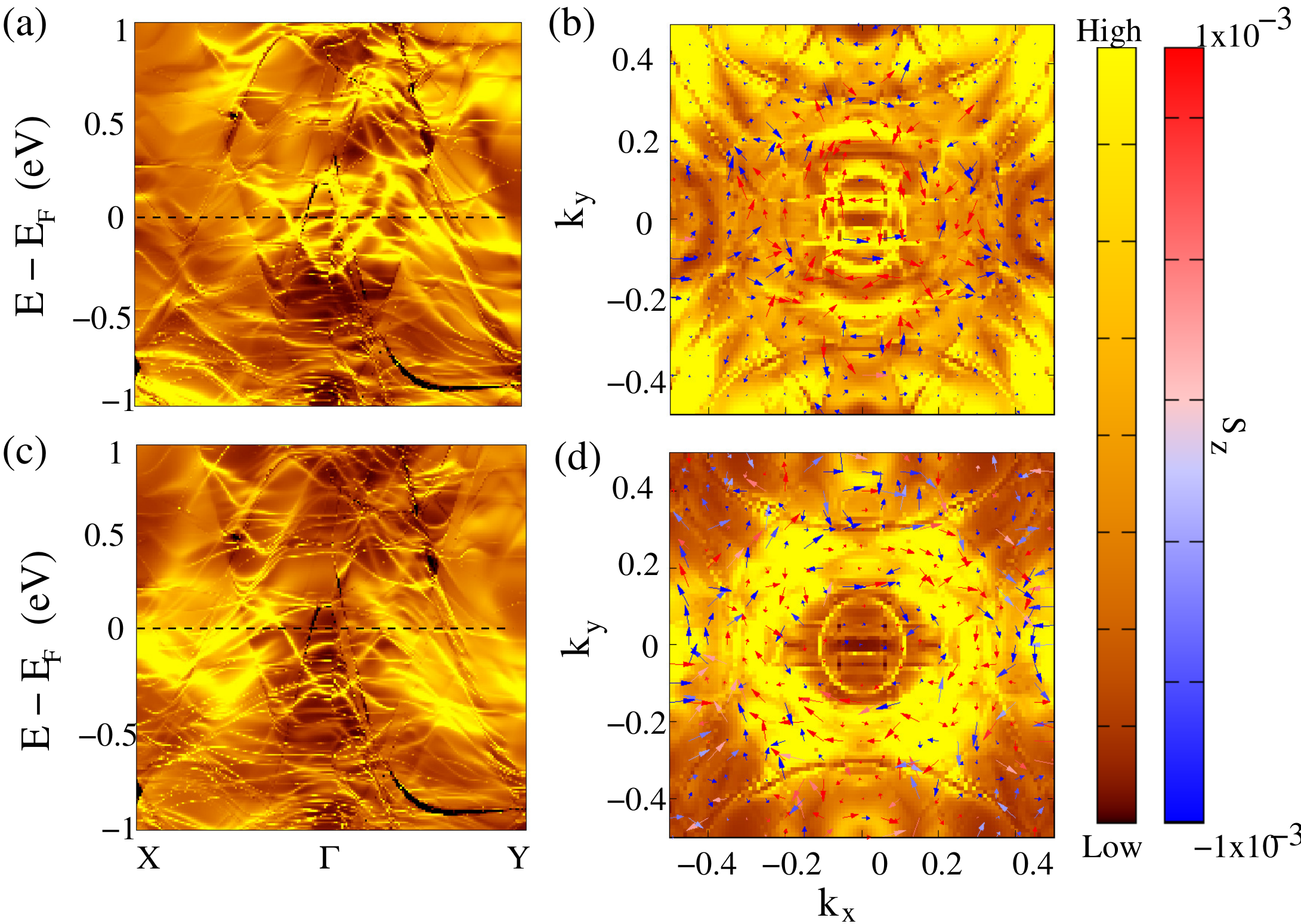} \includegraphics[scale=0.40]{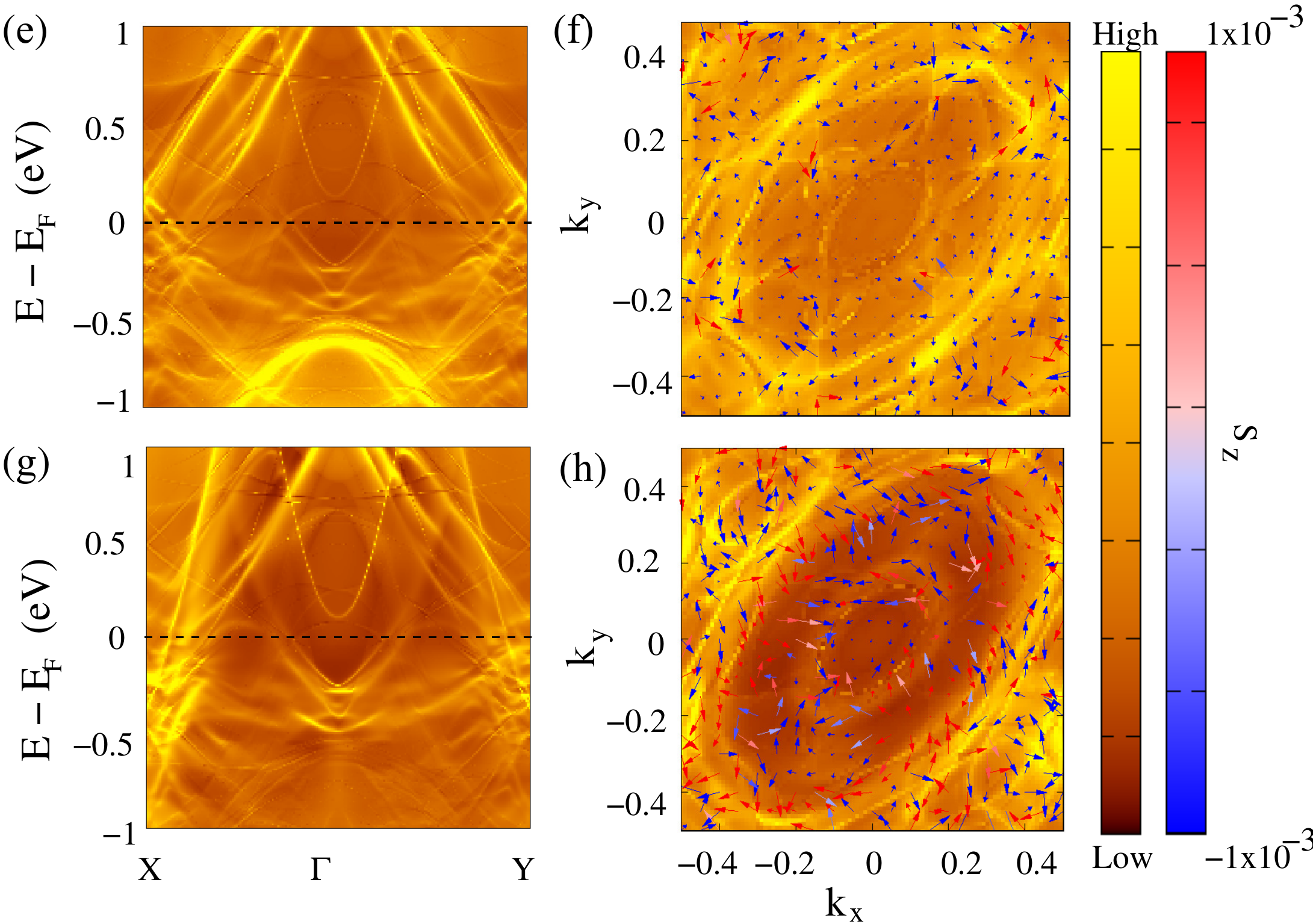} 
	\includegraphics[scale=0.40]{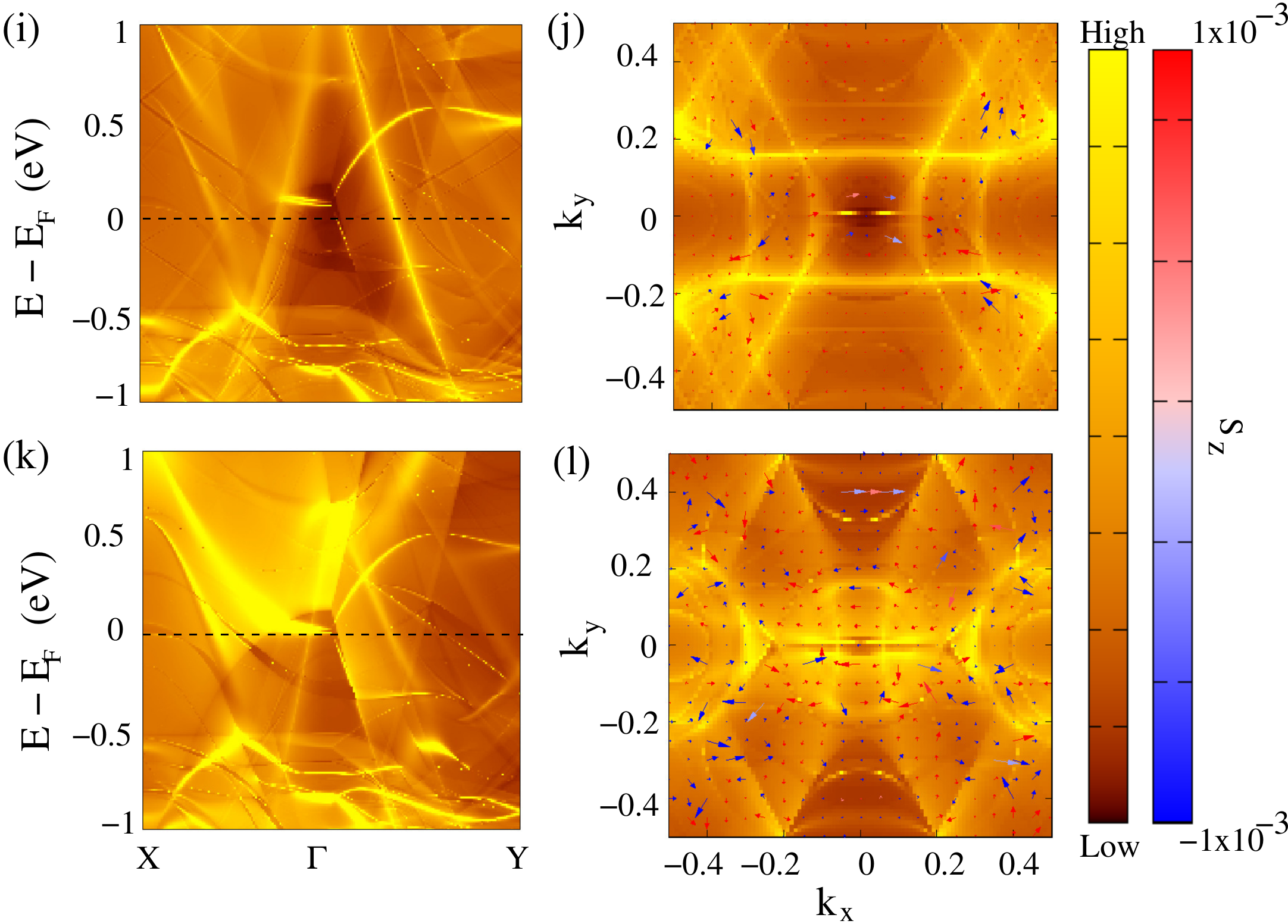} \includegraphics[scale=0.40]{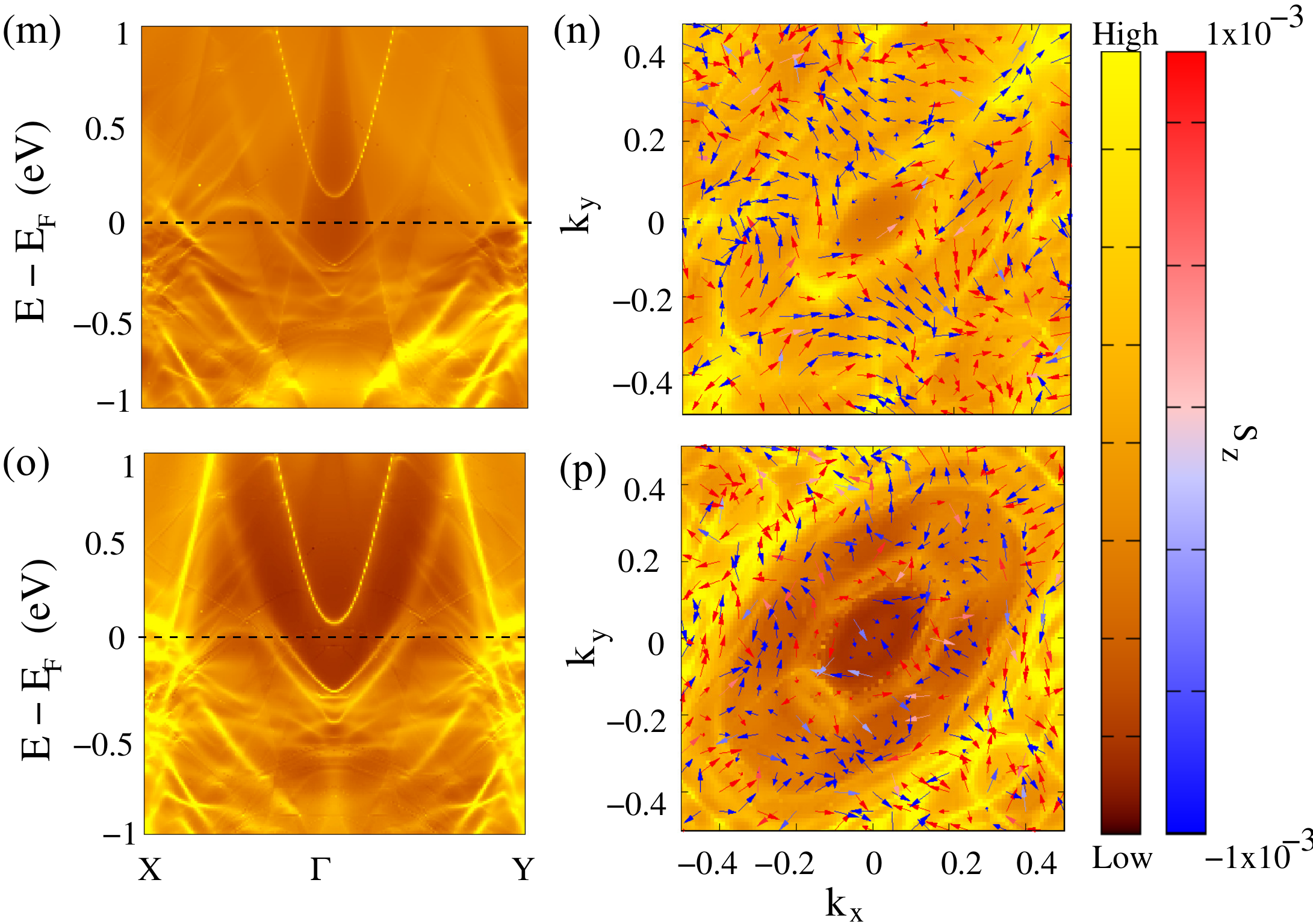} 
	\caption{Spectral function $A(E; k_x, k_y,z\in\{1,2\})$ in Eq.~\eqref{eq:spectral} plotted along high symmetry $k$-path, $X$-$\Gamma$-$Y$ at: (a) plane $1$ in Fig.~\ref{fig:fig1} passing through ML of Co; and (c) plane $2$ in Fig.~\ref{fig:fig1} passing through ML of Ta within Cu/Co/Ta(4 ML)/Cu junction with $\mathbf{m}_\mathrm{Co} || \hat z$. Panels (b) and (d) plot constant energy contours of \mbox{$A(E-E_F^0 =0;k_x,k_y)$} on planes $1$ and $2$, respectively, and the corresponding spin textures where the out-of-plane $S_z$ component of spin is indicted in color (red for positive and blue for negative). Panels (e)--(h) show the same information on planes $1$ and $2$ passing through ML of Co and ML of Pt, respectively, within Co/Pt(4 ML)/Cu junction. Panels (i)--(l) show the same information on planes $1$ and $2$ passing through ML of Co and ML of Cu, respectively, within semi-infinite-Co/semi-infinite-Cu junction. Panels (m)--(p) show the same information on planes $3$ and $4$ in Fig.~\ref{fig:fig1} passing through ML of Pt and ML of Cu, respectively, within Co/Pt(4 ML)/Cu junction. The units for $k_x$ and $k_y$ are 2$\pi/a$ and 2$\pi/b$ where $a$ and $b$ are the lattice constants along the $x$- and the $y$-axis, respectively. The horizontal dashed black line in the spectral function panels denotes the position of the Fermi energy $E_F$.}
	\label{fig:fig3}
\end{figure*}

The abrupt change of $\mathbf{P}$ at different FM/HM interfaces can be read off from the abscissa of Fig.~\ref{fig:fig2} as we change from 0 or 1 ML of HM.  The corresponding values of  parameter $\zeta$ defined in Eq.~\eqref{eq:zeta} are given in Table~\ref{tab:zeta}. Values at 2--5 ML of HM can also be viewed~\cite{Nikolic2005} as the change of $\mathbf{P}$ along HM layer of 5 ML thickness. An infinitely long Co layer generates $|\mathbf{P}| \simeq 0.26$, which is denoted by dashed horizontal line in Fig.~\ref{fig:fig2}(c). Upon attaching Cu to Co layer, and in the absence of any HM interlayer in between them, $|\mathbf{P}|$ reduces to $\simeq 0.20$ in Fig.~\ref{fig:fig2}(a) for Cu/Co(4 ML)/Cu junction or to $|\mathbf{P}| \simeq 0.18$ in  Figs.~\ref{fig:fig2}(c) and ~\ref{fig:fig2}(d) for semi-infinite-Co/semi-infinite-Cu junction. At first sight, it might look surprising that finite thickness \mbox{Co(4 ML)} layer generates larger spin-polarization of charge current than semi-infinite Co layer. However, we note that surfaces of Co (see Fig.~4(a)--(d) in Ref.~\onlinecite{Marmolejo-Tejada2017}) and Cu (see Fig.~5(b) in Ref.~\onlinecite{Ishida2014a}) can  host Rashba SOC enabled by inversion symmetry breaking  where an electrostatic potential gradient can be created by the charge distribution at the metal/vacuum interface to confine wave functions into a spin-split quasi-2D electron gas.~\cite{Bahramy2012}

Since Cu/Co(4 ML)/Cu junction is inversion symmetric, we expect zero SOC at its interfaces~\cite{Miron2010}, while nonzero SOC should appear at the single interface of Co/Cu junction. This is confirmed by plotting spectral function $A(E;k_x,k_y,z)$ at ML of Co on the left side of the interface in Fig.~\ref{fig:fig3}(i), as well as the corresponding spin texture along constant energy contours of the spectral function at  $E-E_F=0$ in Fig.~\ref{fig:fig3}(j). Even larger spin texture is obtained on ML of Cu on the right side of Co/Cu interface which is, therefore, most responsible for SML at Co/Cu interface, thus resolving controversy~\cite{Dassonneville2010} of zero vs. nonzero measured values for $\delta$. Following Ref.~\onlinecite{Marmolejo-Tejada2017}, we obtain the spectral function at an arbitrary plane at position $z$ within the heterostructure in Fig.~\ref{fig:fig1} from the retarded GF in Eq.~\eqref{eq:rgf} 
\begin{equation}\label{eq:spectral}
A(E;k_x,k_y,z)=-\frac{1}{\pi}\mathrm{Im}\,[G_{\mathbf{k}_\parallel}(E;z,z)],
\end{equation} 
where the diagonal matrix elements $G_{\mathbf{k}_\parallel}(E;z,z)$ are computed by transforming the retarded GF from local orbital to a real-space representation. The spin textures within the constant energy contours are computed from the spin-resolved spectral function.

\begin{figure*}
	\includegraphics[scale=0.6]{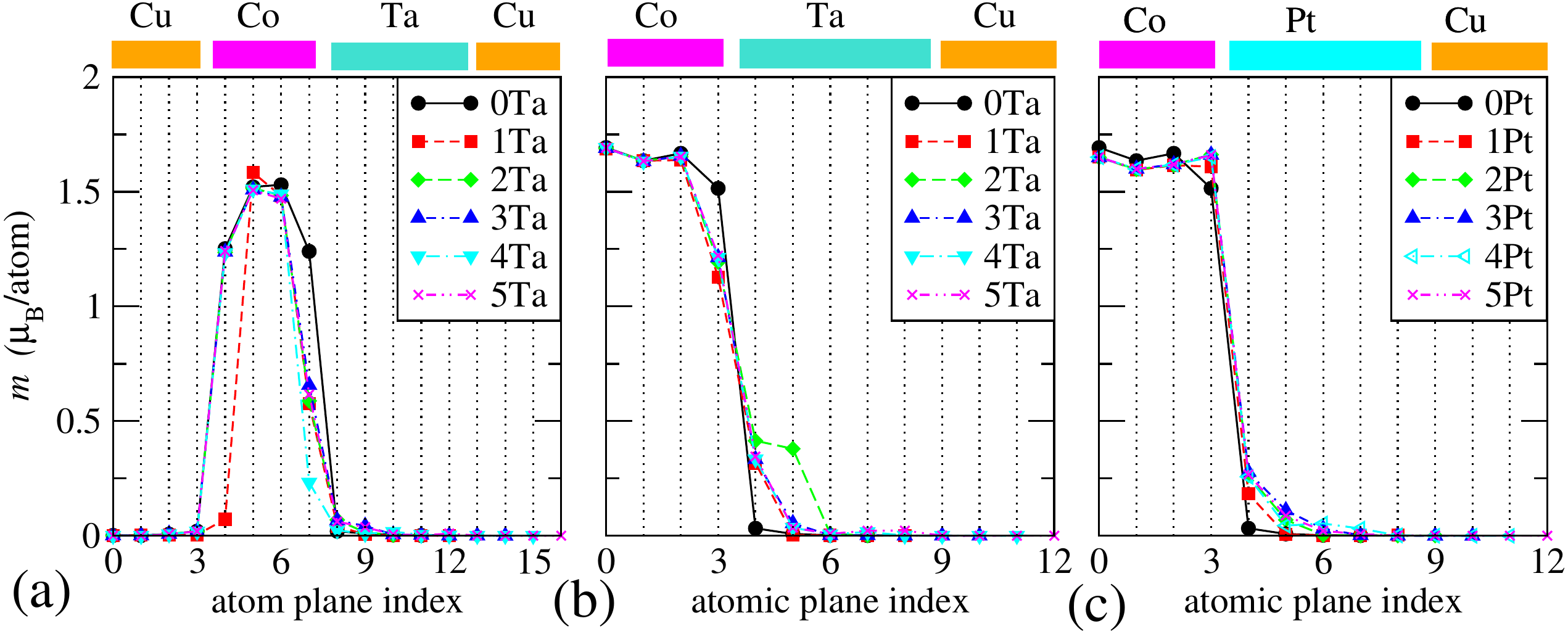}
	\caption{Magnetic moment averaged over the total number of atoms within each ML of: (a) Cu/Co(4 ML)/Ta/Cu, (b) Co/Ta/Cu; and (c) Co/Pt/Cu junctions. The color of symbols and lines denotes different thickness of  Ta or Pt HM layers, ranging from 0 to 5 MLs. Vertical dotted black lines indicate position of atomic MLs. Orange, magenta, and cyan  solid boxes over the top of each panel show spatial extent of Cu, Co, Ta and Pt layers, respectively, as a guide to the eyes.}
	\label{fig:fig4}
\end{figure*}

The parameter $\zeta$ listed in Table~\ref{tab:zeta} is much smaller for Co/Pt interface studied in Fig.~\ref{fig:fig2}(d) than for Co/Ta interface studied in Fig.~\ref{fig:fig2}(a) for the same orientation of magnetization, $\mathbf{m}_\mathrm{Co} \parallel \hat{z}$. Inspection of spectral functions and spin textures around Co/Ta interface in Fig.~\ref{fig:fig2}(a)--(d) and around Co/Pt interface in Fig.~\ref{fig:fig2}(e)--(h) reveals that SML is caused by both Co and Ta MLs in the former case, while it is caused mostly by Pt ML in the latter case. We emphasize that spectral function and spin textures at Co/Pt interface are quite different from those of the ferromagnetic Rashba Hamiltonian~\onlinecite{Nagaosa2010} often employed in calculations of SOT~\cite{Haney2013} or spin pumping in the presence of interfacial  SOC.~\cite{Mahfouzi2012,Chen2015,Shen2014} 

Figure~\ref{fig:fig2} also reveals that while 1 ML of HM always acts as a``spin sink,'' 2 MLs can also act as a ``spin source''~\cite{Amin2016} by re-polarizing charge current flowing from first to second ML of HM. Nevertheless, such ``re-polarization'' is eventually lost sufficiently further away from the FM/HM interface. In the presence of scattering from disorder and/or phonons excited at finite temperature~\cite{Liu2015} within the HM layer, the decay of $\mathbf{P}$ away from the interface would be monotone decreasing exponential function. On the other hand, it has been shown~\cite{Liu2014a} that SML (i.e., the value of $\delta$ or $\zeta$ parameters) can be insensitive to interfacial disorder, despite its strong effect on interface resistances, thereby emphasizing the need to understand SML in clean system in the focus of our study.

Since SML is particularly strong  in Co/Pt/Cu junction in Fig.~\ref{fig:fig2}(d), with $\zeta \simeq 0.04$ signifying almost complete absorption of spin current by the interface, we also investigate  Co/Pt/Au junction in Fig.~\ref{fig:fig2}(d) to find much larger $\zeta \simeq 0.29$. Thus, this difference explains experimentally observed~\cite{Bass2007} large SML at Pt/Cu interface, which is often~\cite{Bass2007,Rojas-Sanchez2014,Lee2009b} naively attributed to interfacial diffusion and disorder. Instead, by plotting spectral functions and spin textures in Fig.~\ref{fig:fig2}(m)--(p) at the MLs of Pt and Cu, where spin textures on both sides of Pt/Cu interface are the largest among the cases in Fig.~\ref{fig:fig2}, we demonstrate that large SML occurs even at {\em perfect} Pt/Cu interface because of strong interfacial SOC. 

In the presence of interfacial SOC electrons trade angular momentum with the atomic lattice that can lead to changes of all three~\cite{Amin2016} components of $\mathbf{P}$. Figure~\ref{fig:fig2}(b) shows that for injected spins polarized parallel to the interface---i.e., $P_x \neq 0$ while the other two components of the incoming  polarization vector are zero due to $\mathbf{m}_\mathrm{Co} \parallel \hat{x}$---SML reduces the magnitude of $P_x$ while also generating non-negligible $P_z$ out of the plane. Interestingly, very recent experiments~\cite{Baek2017} on lateral FM1/NM/FM2 trilayers with charge current injected parallel to the plane, where magnetization of FM1 layer is in-plane and fixed while the magnetization of FM2 layer is out-of-plane and free, have observed out-of-plane spin current with nonzero $P_z$ as advantageous for switching magnetization of FM2 layer via torque in the absence of any externally applied magnetic field. This is fully compatible with Fig.~\ref{fig:fig2}(b), once we take into account that spin current impinging perpendicularly onto FM1/NM interface originates from the anomalous Hall effect~\cite{Nagaosa2010} in the FM1 layer and carries only in-plane polarized spins.

Finally, since change of polarization vector along HM layer of different thicknesses in Fig.~\ref{fig:fig2}, as well as peculiar thickness dependence of SOT observed in Co/Ta bilayers~\cite{Kim2013}, can be affected by the magnetic proximity effect~\cite{Lim2013,Grytsyuk2016}---where FM induces nonzero magnetic moments within MLs of HM sufficiently close to the interface---we plot in Fig.~\ref{fig:fig4} spatial profile of average magnetic moment per atom across junctions studied in Fig.~\ref{fig:fig2}. The characteristic magnetization decay length within Pt in Fig.~\ref{fig:fig4}(c) is longer than in Ta in Figs.~\ref{fig:fig4}(a) and ~\ref{fig:fig4}(b), where 1 ML of Ta is also capable of surprisingly large suppression of magnetic moments on the FM side for both finite and infinite thickness of the Co layer.

\begin{acknowledgments}
We thank K. D.  Belashchenko, A. A. Kovalev, K.-J. Lee and Z. Yuan for insightful discussions. This work was supported by DOE Grant No. DE-SC0016380. The supercomputing time was provided by XSEDE, which is supported by NSF Grant No. ACI-1053575. 
\end{acknowledgments}


\end{document}